\documentclass[showpacs,prl,onecolumn,aps,superscriptaddress,preprintnumbers,nofootinbib]{revtex4}
\usepackage[T1]{fontenc}
\usepackage[latin9]{inputenc}
\usepackage{amsmath,amssymb}
\usepackage{epsfig}
\usepackage{graphicx}
\usepackage{mathrsfs}
\usepackage{amsmath}
\usepackage{amsfonts}
\usepackage{epstopdf}
\usepackage{color}
\def\slashchar#1{\setbox0=\hbox{$#1$}     		
   \dimen0=\wd0                                 	
   \setbox1=\hbox{/} \dimen1=\wd1               	
   \ifdim\dimen0>\dimen1                        	
      \rlap{\hbox to \dimen0{\hfil/\hfil}}      	
      #1                                        	
   \else                                        	
      \rlap{\hbox to \dimen1{\hfil$#1$\hfil}}   	
      /                                         	
   \fi}

\newcommand{\beq}{\begin{equation}}
\newcommand{\eeq}{\end{equation}}
\newcommand{\bea}{\begin{eqnarray}}
\newcommand{\eea}{\end{eqnarray}}
\newcommand{\ba}{\begin{array}}
\newcommand{\ea}{\end{array}}

\def\eq#1{{Eq.~(\ref{#1})}}
\def\fig#1{{Fig.~\ref{#1}}}
\newcommand{\bas}{\bar{\alpha}_S}

\newcommand{\nn}{\nonumber}

\newcommand{\Lb}{\left(}
\newcommand{\Rb}{\right)}

\newcommand{\pom}{I\!\!P}
\newcommand{\Y}{\tilde{Y}}

\begin{document}

\title{Can $1/N_c$ corrections destroy the saturation of dipole densities?}

\author{Eugene Levin}
\email{leving@tauex.tau.ac.il, eugeny.levin@usm.cl}
\affiliation{Department of Particle Physics, Tel Aviv University, Tel Aviv 69978, Israel}
\affiliation{Departemento de F\'isica, Universidad T\'ecnica Federico Santa Mar\'ia, and Centro Cient\'ifico-\\
Tecnol\'ogico de Valpara\'iso, Avda. Espana 1680, Casilla 110-V, Valpara\'iso, Chile}

\date{\today}

\pacs{13.60.Hb, 12.38.Cy}

\begin{abstract}
In this paper we discuss a well known QCD result: the steep increase of  Green's function for exchange of $n$  BFKL Pomerons  $G_{n \pom}\Lb Y\Rb\propto \exp\Lb \frac{n^2}{N^2_c} \Delta_{\mbox{\tiny BFKL}} Y\Rb$ where $N_c $ is the number of colours ,Y is the rapidity and $\Delta_{\mbox{\tiny BFKL}}$ is the intercept of the BFKL Pomeron.
We consider this problem in the framework  of the simple  Pomeon models in zero transverse dimensions, which have two advantages :(i)  they  allow to take into account all shadowing corrections, including the summation of the Pomeron loops and (ii) they have the same as in QCD striking increase of $G_{n \pom}\Lb Y\Rb $. We found that the strength of shadowing corrections is not enough to stop the increase of the scattering amplitude with energy in contradiction to the unitarity constraints. Hence, our answer to the question in the title is positive. We believe that we need to search an approach beyond of the BFKL Pomeron calculus to treat $1/N_c$ corrections in Colour Glass Condensate effective theory.

 \end{abstract}
\maketitle
\vspace{-0.5cm}
\tableofcontents


\section{Introduction}

The only candidate for the effective theory of high energy QCD is Colour Glass Condensate(CGC) approach
(see  Ref.\cite{KOLEB} for a review). Two main ideas of CGC: 
the saturation of the dipole density and the new dimensional scale ($Q_s$), which increases with energy, have become  widely accepted language for discussing the high energy scattering in QCD. However the CGC approach suffers several problems. The most known of them is the power-like behaviour of the scattering amplitude at large impact parameters\cite{KW1,KW2,KW3,FIIM} that violated the Froissart theorem\cite{FROI}. We have to introduce non-perturbative corrections at large impact parameters and the embryonic stage of our understanding of confinement of quarks and gluons does not allow us to come up with a reasonable theoretical approach to  the problem.  The second well known problem is summing of the BFKL Pomeron loops\footnote{ The abbreviation BFKL Pomeron  stands  for Balitsky, Fadin, Kuraev and Lipatov Pomeron.}. This problem is a technical one, but in spite of intensive work 
 \cite{BFKL,LIP,LIREV,LIFT,GLR,GLR1,MUQI,MUDI,Salam,NAPE,BART,BKP,MV,MUSA,KOLE,BRN,BRAUN,BK,KOLU,JIMWLK1,JIMWLK2,JIMWLK3,JIMWLK4,JIMWLK5,JIMWLK6,JIMWLK7,JIMWLK8,AKLL,KOLU1,KOLUD,BA05,SMITH,KLW,KLLL1,KLLL2,LEN}
 it is still far away from being solved.This situation makes the problem  one of the  principle problems,  without solving which we cannot consider the dilute-dilute and dense-dense parton densities collisions. As has been recently shown\cite{KLLL1,KLLL2}, even the Balitsky-Kovchegov (BK) equation, that governs the dilute-dense parton density  scattering (deep inelastic scattering (DIS) of electron with proton), has to be modified due to  contributions of  Pomeron loops.
  
\begin{figure}[ht]
    \centering
  \leavevmode
      \includegraphics[width=9cm]{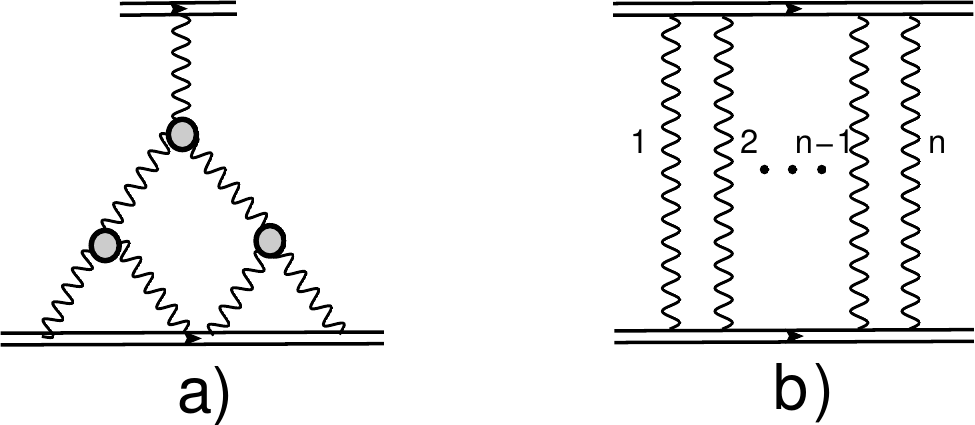}  
      \caption{\fig{fan}-a:  the `fan' diagrams of the Balitsky-Kovchegov non-linear equation. The wavy lines denote the BFKL Pomerons. The blob shows the triple Pomeron vertex. \fig{fan}b: is the Green's function of the exchange of $n$ BFKL Pomerons.}
\label{fan}
   \end{figure}

  In this paper we wish to draw an attention of the reader to a different problem of the CGC approach. In the CGC approach the scattering amplitude does not exceed the unitarity limit due to shadowing corrections. The Balitsky-Kovchegov \cite{BK} non-linear equation, which sums the `fan' diagrams of interacting BFKL Pomerons (see \fig{fan}-a), generates the amplitude, which tends to unity (the unitary limit) at high energies.  However, it has been shown in Refs.\cite{LRS,BART0,BARTU,LLS,LLR}
  that  Green's function of the exchange of $n$ BFKL Pomerons does not increase as $\exp\Lb n\,\Delta_{\mbox{\tiny BFKL}}Y\Rb$ where $\Delta_{\mbox{\tiny BFKL}}$   is the intercept of the BFKL Pomeron and $Y$ is the rapidity of two colliding dipoles. It turns out that  Green's function grows as
   $\exp\Lb \frac{n^2}{N^2_c}\Delta_{\mbox{\tiny BFKL}}Y\Rb$ ($N_c$ is the number of colours)  and this increase cannot be suppressed by the shadowing corrections
   \footnote{ Actually,  in  Refs.\cite{LRS,BART0,BARTU,LLS,LLR}   the exchange of  $n$-BFKL Pomerons were considered in the double log  approximation, which leads to the anomalous dimension $\frac{\bas}{(N^2-1)^2\omega} n^4$ instead of   $\frac{\bas}{\omega} n^2$ for the exchange of $n$ BFKL Pomerons .}. The $N_c$ suppression gives rise for the hope that  such an increase can manifest itself only at very high energies, but recently  A. Kovner and  M. Li showed that for nucleus-nucleus scattering in the CGC approach there exist  large $1/N^2_c$  corrections that  are larger than the shadowing suppression\footnote{ We thank A, Kovner and M. Li for sharing with us their finding.}. In spite of these dangerous    results for the CGC approach, the references\cite{LRS,BART0,BARTU,LLS,LLR} 
    show that these large corrections can be treated in the BFKL Pomeron calculus if we introduce the vertices of interaction of four Pomerons ($\pom + \pom \to \pom + \pom$).

  Bearing this in mind we wish to return to discussion of these corrections  in 
 simple, but exactly solvable, two dimensional models\cite{MUSA,ACJ,AAJ,JEN,ABMC,CLR,CIAF,BIT,RS,KLremark2,SHXI,KOLEV,nestor,LEPRI,utm,utmm}.   In these models we can sum Pomeron loops and we have the same kind of corrections, which go under slang name  of many particle Regge poles      \cite{DISASTER}. In the next section we will discuss these corrections in details. Here we wish to tell that the main goal of this paper is to find the scattering amplitude taking into account the $\pom + \pom \to \pom + \pom$ vertices   in the simple two dimensional models to get experience what we can expect in QCD in the CGC approach for these  disastrous contributions.

\section{Setting the problem}

\begin{figure}[ht]
    \centering
    \begin{tabular}{c c}
  \leavevmode
      \includegraphics[width=14cm]{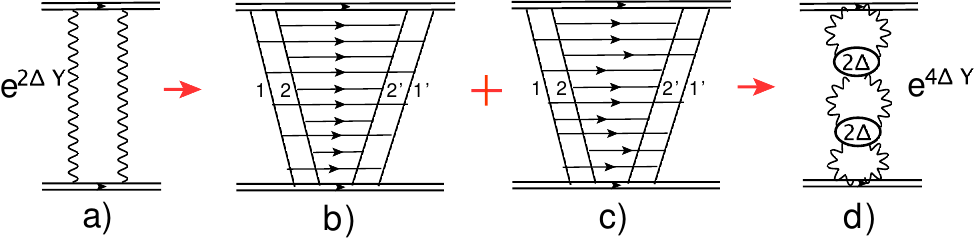}& \includegraphics[width=3.9cm]{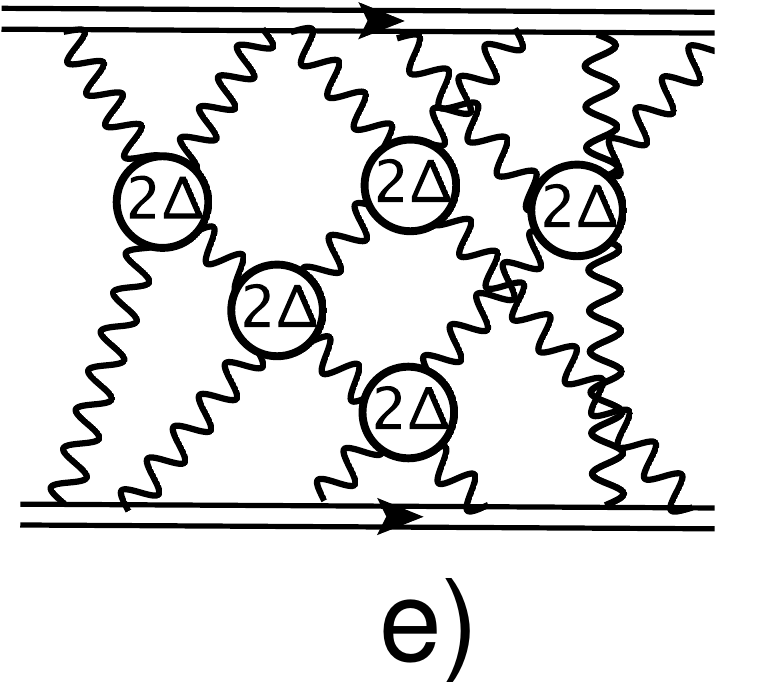}\\
      \end{tabular}   
         \caption{ Summing the switch diagrams for the exchange of two BFKL Pomerons.
      \fig{2pom}-a: the exchange of two BFKL Pomerons. \fig{2pom}-b: the exchange of two BFKL Pomeron in the two dimensional model.  \fig{2pom}-c: the switch diagram for two ladder exchange in which the produced dipoles from the Pomeron 1 are absorbed by the Pomeron 2' and vise versa.  \fig{2pom}-d: the sum of all diagrams in the two dimensional model that contribute in the Green's  function of the exchange of two Pomerons in t-channel.  \fig{2pom}-e: the Green's function of the exchange of n$n$ Pomerons.The wavy lines denote the exchange of one Pomeron with the Green's function $\exp\Lb \Delta Y\Rb$. The $\pom + \pom \to \pom + \pom$ vertex is denoted by blob and equals to $2 \Delta$.}
\label{2pom}
   \end{figure}

 Green's function for one Pomeron in the two dimensional model  can be viewed as a sum of the contributions:
   \beq   \label{POM1}
 G_{\pom}\Lb Y\Rb \,\,=\,\,\sum_{k=0}^\infty\,\frac{\Lb \Delta\,Y\Rb^n}{n!}\,\,=\,\,e^{\Delta\,Y}
 \eeq  
 \eq{POM1} sums the ladder diagrams in leading log(1/x) approximation in which 
all produced dipoles have a strong ordering in the fractions of total momentum $x_i$:
  \beq \label{POM2}
  1\,\gg\,x_1\,\gg\,x_2  \,\gg\, \dots \,x_i\,\gg\,x_{i+1} \,\gg\,\dots\,\gg\,\,x_{n-1}\,\gg\,x_n
  \eeq
this leads to $1/n!$ in \eq{POM1}. The two Pomeron exchange is shown in \fig{2pom}-a and  corresponds to the Feynman diagram of \fig{2pom}-b in which all dipoles emitted by the ladder 1 are absorbed by ladder 1', and all  dipoles produced  by ladder 2 are absorbed by ladder 2'. These diagrams lead to the Green's function of the exchange of two Pomerons: $G_{2 \pom} =G^2_{\pom}\Lb Y\Rb $.  However, the diagrams in which  the dipole,  emitted from ladder 1, will be absorbed by ladder  2'  are not small in the two dimensional models. We are going to call these diagrams  ``switch diagrams" , using the terminology suggested in Refs.\cite{LRS,BART0,BARTU,LLS,LLR}.  Indeed, after first exchange ladder 1 and 2' will give the Pomeron exchange, leading to the diagram of \fig{2pom}-d. Since at given rapidity we have two `switch diagrams: dipole emitted between ladders 1 and 2' and the dipole emitted between ladders 2 and 1' we can  obtain the vertex $\pom + \pom \to \pom + \pom$ , which is equal to $ 2 \Delta$.

The diagrams of \fig{2pom}-d can be summed  in $\omega $-representation:

\beq \label{POM3}
G_2\Lb \omega \Rb\,\,=\,\,\sum_{k=0}^\infty \frac{1}{\omega\,-\,2\,\Delta} \Lb \frac{2\,\Delta}{\omega\,-\,2\,\Delta}\Rb^k\,=\,\frac{1}{\omega\,-\,4 \,\Delta}
\eeq
Where $1/\Lb \omega - 2 \Delta\Rb$ is the contribution of the exchange of two Pomerons (see \fig{2pom}-b).  Coming back to $Y$ representation one can see that $G_2\Lb Y\Rb\,\,=\,\,\exp\Lb 4\,\Delta\,Y\Rb$ instead of $G_2\Lb Y\Rb\,\,=\,\,\exp\Lb 2\,\Delta\,Y\Rb$  which is expected for two Pomeron exchange.
Summing the diagrams of \fig{2pom}-e we obtain the Green's function for the exchange of $n$ Pomerons: $G_n\Lb Y\Rb\,=\,\,\exp\Lb n^2 \Delta Y\Rb$. This Green's function was derived in Ref.\cite{DISASTER} for the parton approach to high energy scattering. In QCD the structure of the corrections remain to be the same and the only difference  is that the `switch' diagrams as well the intercept of $n$ Pomeron state have the smallness of the order of $1/N^2_c$. Therefore, the $\pom + \pom \to \pom + \pom$ vertex in QCD is equal to $\frac{2}{N^2_c}\,\Delta_{\mbox{\tiny BFKL}}$. leading to
$G_n\Lb Y\Rb\,=\,\,\exp\Lb n \Delta\,Y\,+\, \frac{n(n - 1)}{N^2_c} \Delta Y\Rb$.

\section{ The structure of the parton cascade}

The useful tool  for discussing  the structure of the parton cascade is the generating function, which has the following form\cite{MUDI,LELU}:
\beq \label{Z}
Z\Lb Y, u\Rb\,\,=\,\,\sum_{n}P_n(Y) \,u^n
\eeq
where $P_n(Y) $ is the probability to find $n$ dipoles with rapidity $Y$. For the scattering of one dipole  with the target we have the following initial and boundary conditions:
\beq \label{IC}
\mbox{initial condition:}\,\, Z\Lb Y=0,u\Rb = u; ~~~~~\mbox{boundary  condition:}\,\, Z\Lb Y,u=1\Rb\,=\,1;
\eeq
The boundary condition follows from $P_n(Y)$ being probabilities.

We need to write the evolution equation for $Z\Lb Y, u\Rb$ taking into account  two Pometon vertices: $\Gamma_{3 \pom} = \Delta $ for $\pom \to \pom + \pom$ and $\Gamma^{2 \pom}_{2 \pom}  \,=\,2 \,\Delta $ for $\pom  + \pom \to \pom + \pom$. Fortunately  the equation for $Z$ for such a cascade has been written in Ref.\cite{LELU} and it takes the form:
\beq \label{ZEQ}
\frac{ \partial\,Z\Lb Y,u\Rb}{\partial Y}\,\,=\,\,- \Delta u (1-u) \frac{ \partial\,Z\Lb Y,u\Rb}{\partial u}\,\,+\,\,\Delta\,u (u - 1)  \frac{ \partial^2\,Z\Lb Y,u\Rb}{\partial u^2}
\eeq
As has been mentioned that \eq{ZEQ} has been suggested in Ref.\cite{LELU} , but \eq{ZEQ}  has a remarkable difference in comparison with equations in this reference: the sign in front of $u^2 \partial^2/\partial\,u^2$ is plus, not minus as in our previous attempts.

 Below we discuss a bit different equation:
 \beq \label{ZEQ1}
\frac{ \partial\,Z\Lb \Y,u\Rb}{\partial \Y}\,\,=\,\,- \ u (1-u) \frac{ \partial\,Z\Lb \Y,u\Rb}{\partial u}\,\,+\,\,\kappa\,u (u - 1)  \frac{ \partial^2\,Z\Lb \Y,u\Rb}{\partial u^2}
\eeq 
where $\Y = \Delta\,Y$ and factor $\kappa = 1/N^2_c$ takes into account a suppression for $\Gamma^{2 \pom}_{2\pom} = 1/N^2_c$  in QCD. For two dimensional models $\kappa = 1$.

\subsection{BFKL cascade}

Neglecting $\Gamma^{2 \pom}_{2 \pom}$ we obtain the well known equation for the BFKL cascade.
 \beq \label{ZEQ2}
\frac{ \partial\,Z\Lb \Y,u\Rb}{\partial \Y}\,\,=\,\,- \ u (1-u) \frac{ \partial\,Z\Lb \Y,u\Rb}{\partial u}
\eeq 
It is instructive to observe that \eq{ZEQ2}  leads to  a non-linear equation for $ Z\Lb Y, u\Rb$\cite{MUDI,LELU}.   Indeed, the general solution to \eq{ZEQ2}  is of the form  $Z (Y, u) = Z(u(Y ))$; if we substitute this function into \eq{ZEQ2} , the derivatives  $\partial Z/\partial Y$ on the l.h.s. and r.h.s. of \eq{ZEQ2} cancel, and we obtain a differential equation for the function $u(Y)$.  Using the initial condition of \eq{IC} we can re-write
\eq{ZEQ2} in the form:
\beq \label{ZEQ3}
\frac{ \partial\,Z\Lb \Y,u\Rb}{\partial \Y}\,\,=\,\,- \, Z\Lb Y, u\Rb  \,\,+\,\,Z^2\Lb Y, u\Rb
\eeq
Note, that the scattering amplitude in our models is equal to $N\Lb Y\Rb\,\,=\,\,1\,\,-\,\,Z\Lb Y,1-\gamma\Rb$ where $\gamma$ is the amplitude of the interaction of the dipole with the target at low energy.  For $N\Lb Y\Rb$ we  have the nonlinear equation:
\beq \label{ZEQ4}
\frac{ \partial\,N\Lb \Y\Rb}{\partial \Y}\,\,=\,\,- \, N\Lb Y\Rb  \,\,+\,\,N^2\Lb Y\Rb
\eeq
which is the Balitsky-Kovchegov equation\cite{BK}  for our simple models.

The solution to \eq{ZEQ3} , which satisfies the initial and boundary conditions of \eq{IC} has the following form:
\beq \label{ZEQ5}
Z\Lb Y, u\Rb\,\,=\,\,\frac{u\, e^{-\Delta \,Y}}{1\,\,+\,\,u\Lb e^{-\Delta \,Y}\,\,-\,\,1\Rb}
\eeq
One can see that at large values of $Y$  $Z \,\to\,0 $  ($N \,\to\,1$). In other words, the nonlinear corrections, which stem from  triple Pomeron interactions , suppress the increase of the scattering amplitude ($N \propto\,e^{\Delta\,Y}$) and lead to the scattering amplitude which reaches the unitarity bound. We call this phenomenon the saturation of parton densities.

\subsection{Asymptotic solution}

 The general solution to \eq{ZEQ2}   has been found in Ref.\cite{KOLEV}. However, before  applying the developed technique to this particular equation we wish to see a qualitative changes in the behaviour of $Z$ at large values of $Y$ that stems from a different sign of $\Gamma^{2 \pom}_{2 \pom}$ than in previous attempts to develop a similar cascade.

For this purpose we are going to find the asymptotic solution at large $Y$ from the following equation:
\beq \label{5}
\frac{\partial\,Z^{\rm asymp}\Lb \Y,u\Rb}{\partial u}\,\,+\,\,\kappa  \frac{ \partial^2\,Z^{\rm asymp}\Lb \Y,u\Rb}{\partial u^2}\,\,=\,\,0
\eeq
It has an obvious solution 
\beq\label{ZAS}
 Z^{\rm asymp}\Lb u\Rb  =\,\,\frac{1\,\,-\,\, e^{-\frac{ u}{\kappa}}}{1\,\,-\,\, e^{-\frac{ 1}{\kappa}}},
 \eeq 
which satisfies the boundary condition : $Z^{\rm asymp}\Lb  u=1\Rb  \,\,=\,\,1$  and  $Z^{\rm asymp}\Lb u\Rb \propto u $ at $u \ll  1$. 
To find, how the solution approaches the asymptotic one, we  are looking for the solution in the form:   $ Z\Lb Y, u\Rb\,\,=\,\,\Lb 1\,-\,\,e^{ - \frac{u}{\kappa} -\phi(Y,u)}\Rb/\Lb 1\,\,-\,\, e^{-\frac{ 1}{\kappa}}\Rb$, assuming that $\phi''_{uu}$ and $\phi'^2_u$ are small. The equation for $\phi$ takes the form:

\beq \label{6}
\phi'_{\Y}\Lb \Y,u\Rb\,\,=\,\,u\,(1- u)\,\phi'_u\Lb \Y,u\Rb
\eeq
with the initial condition at Y=0:

\beq \label{IC}
\phi\Lb \Y=0,u\Rb\,\,=\,\, \Lb \frac{1}{C}\,-\,\frac{1}{\kappa} \Rb\,u (1 - u) ~~~\mbox{with}~~C= \frac{1}{ 1 - \exp\Lb 1/\kappa\Rb}
\eeq

which corresponds to $Z\Lb Y=0,u\Rb = u$ and $Z\Lb Y=0,u=1\Rb = 1$.  A general solution to \eq{6} has the form:
\beq \label{7}
\phi\Lb Y,u\Rb\,\,=\,\,\Phi\Lb \Y  + \ln \frac{u}{1 - u}\Rb
\eeq
where the arbitrary function $\Phi$ has to be found from the initial conditions of \eq{IC}.

Finally, the solution takes the form:
\beq \label{8}
 Z\Lb Y, u\Rb\,\,=\,\,\frac{1}{1 - e^{-1/\kappa}} \Bigg( 1 - \exp\Lb - \frac{u}{\kappa}-
 \Lb ( 1 - \exp\Lb -1/\kappa\Rb) -  \frac{1}{\kappa} \Rb
 \frac{u(1-u)\,e^{\Y}}{ (1 - u + u e^{\Y})^2} \Rb\Bigg)
\eeq

\begin{figure}[ht]
    \centering
  \leavevmode
  \begin{tabular}{c c} 
      \includegraphics[width=8cm]{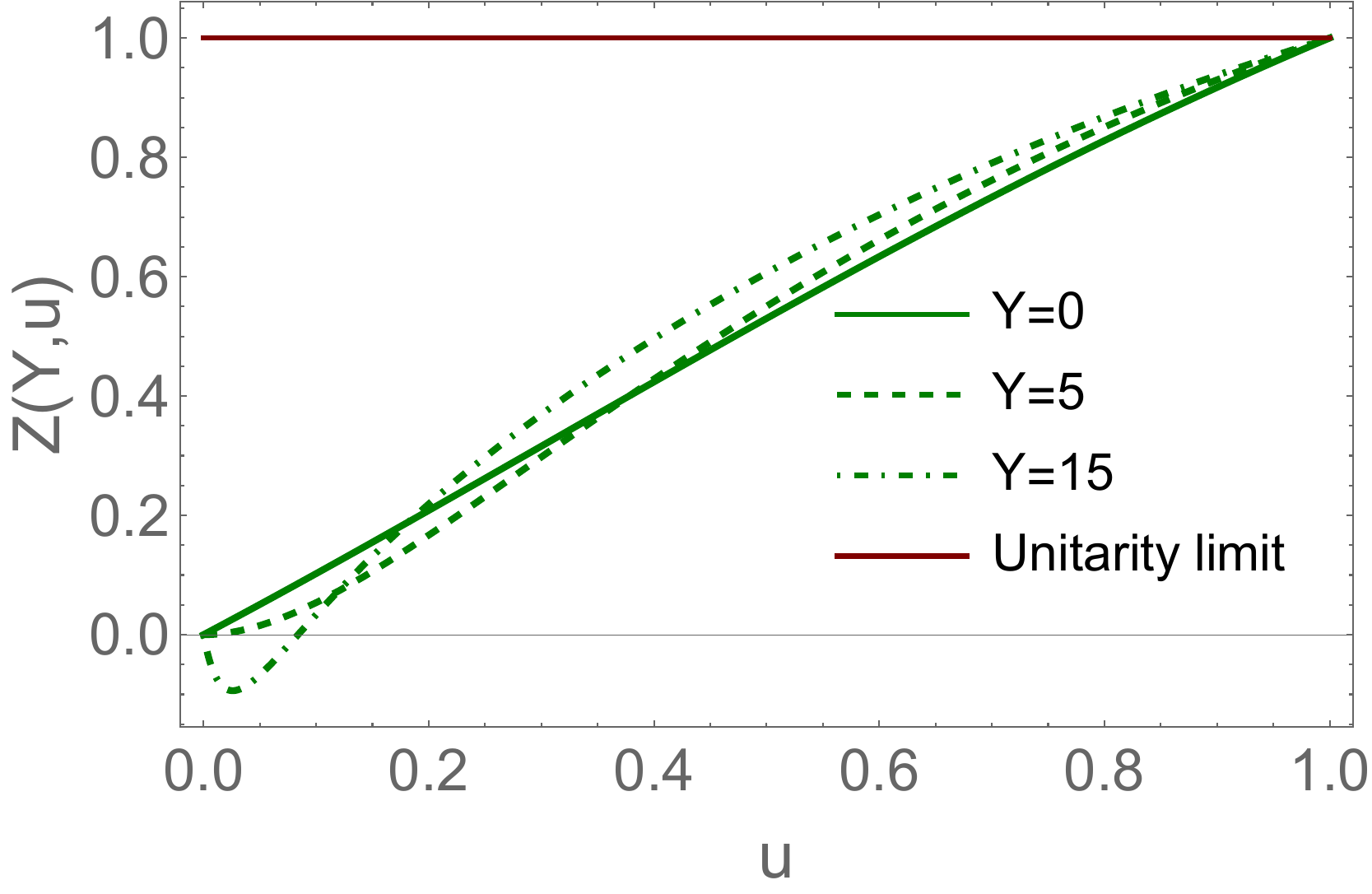} &  \includegraphics[width=8cm]{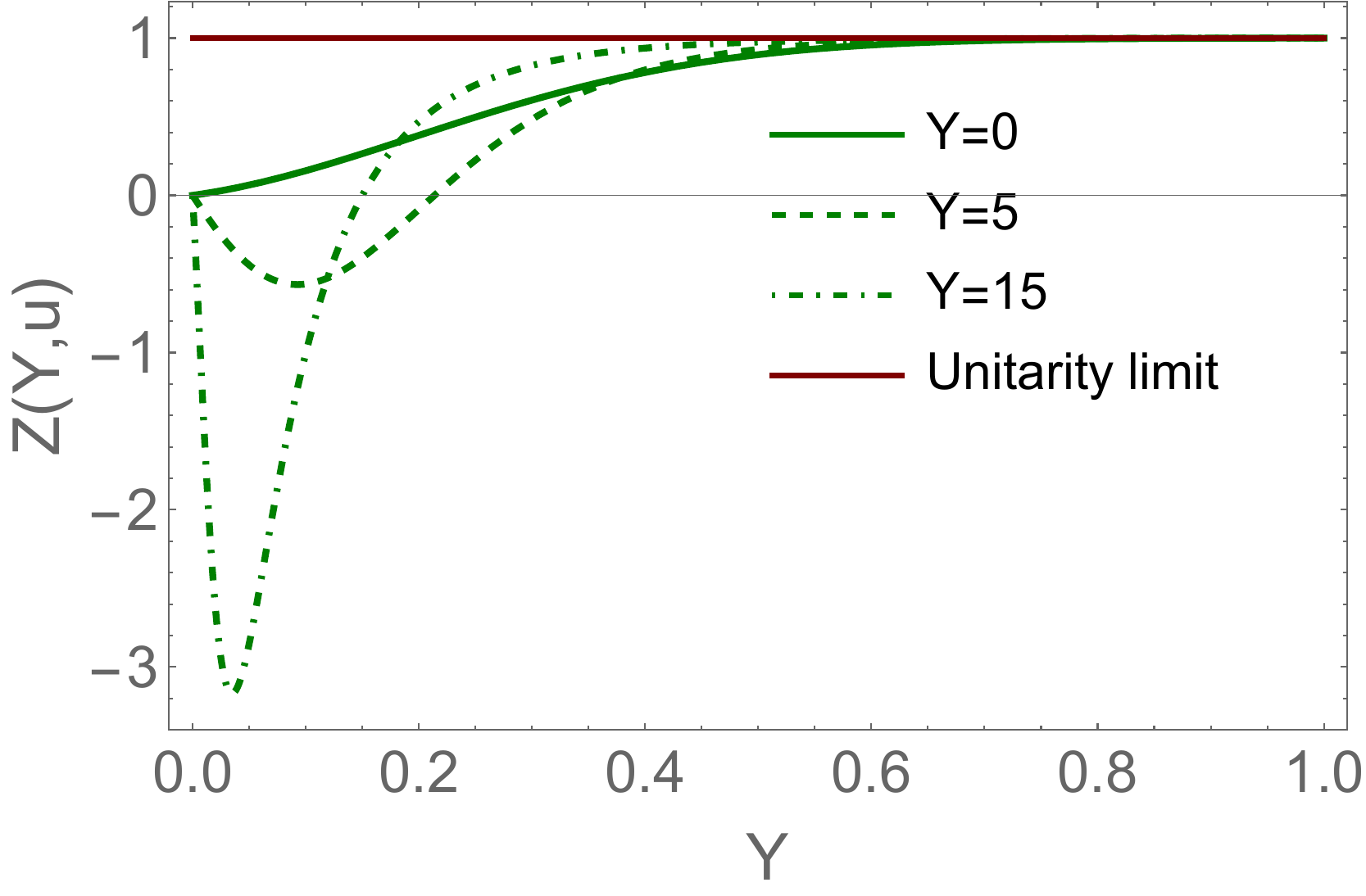}\\
      \fig{zz}-a & \fig{zz}-b\\
      \end{tabular}
      \caption{Graphic form of the solution of \eq{8}.\fig{zz}-a: $\kappa =1$. \fig{zz}-b: $\kappa =1/8$. $\Delta = 0.2$. }
\label{zz}
   \end{figure}

\fig{zz} shows that this solution approaches  the unitarity limit, giving a hope that the shadowing corrections could  suppress the increase of the Green's function of the $n$-Pomerons in $t$-channel. However, this figure shows  that  $Z\Lb \Y,u\Rb  \,<\,0$ in the limited range of $u$, where this solution violates the unitarity constraints.
Hence, we need to find exact solution for the final conclusions. 

\subsection{Exact solution}


Fortunately, the general solution to \eq{ZEQ2} has been found in Ref.\cite{KOLEV}.
We discuss this solution here, repeating all steps of Ref.\cite{KOLEV} and paying special attention to the sign in the r.h.s. of \eq{ZEQ2}. First, we consider $Z\Lb Y, u\Rb$ in $\omega$ representation:

\beq \label{GSOL1}
Z\Lb Y, u\Rb\,\,=\,\,\int\limits^{\epsilon + i \infty}_{\epsilon-i \infty} \frac{d \omega}{2\,\pi\,i} e^{ \omega\,\Y} z\Lb \omega, u\Rb
\eeq
For $z\Lb \omega, u\Rb$ \eq{ZEQ2} takes the form:
\beq \label{GSOL2}
\omega\,z\Lb \omega, u\Rb\,\,=\,\,- u(1-u) \Big(  \frac{\partial\, z\Lb \omega, u\Rb}{\partial\,u}\,\,+\,\,\kappa \frac{\partial^2\, z\Lb \omega, u\Rb}{\partial\,u^2}\Big)
\eeq
Plugging $z\Lb \omega, u\Rb\,\,=\,\,\exp\Lb - \frac{u}{2 \kappa}\Rb {\tilde z}\Lb \omega, u\Rb$ in \eq{GSOL2} we obtain:
\beq \label{GSOL3}
4\,\kappa\,\omega\,{\tilde z}\Lb \omega, u\Rb\,\,=\,\,-u(1-u) \Big( - {\tilde z}\Lb \omega, u\Rb\,\,+\,\,4\kappa^2 \,{\tilde z}''_{uu}\Lb \omega, u\Rb\Big)
\eeq
Introducing $\tilde{z}\Lb \omega, u\Rb= u(1-u) {\cal G}\Lb \omega, u\Rb =  \frac{1 - v^2}{4} {\cal G}\Lb \omega, v\Rb$ with $ 1- 2u = v$, we can rewrite \eq{GSOL3} in the following form:
\bea \label{GSOL4}
4\,\kappa\,\omega{\cal G}\Lb \omega, u\Rb\,\,&=&\,\,u (1 - u){\cal G}\Lb \omega, u\Rb\,\,-\,\,4\kappa^2 \Lb u(1-u)\, {\cal G}\Lb \omega, u\Rb\Rb''_{u u}\nn\\
&=& \Lb u(1-u) + 8\kappa^2\Rb{\cal G}\Lb \omega, u\Rb\,+\,8\,\kappa^2 (2 u - 1) {\cal G}'_u\Lb \omega, u\Rb\,\,-\,\,4\kappa^2 u (1-u){\cal G}''_{uu}\Lb \omega, u\Rb\eea
Using $v$  we have
\beq \label{GSOL5}
(1 - v^2) \,{\cal G}''_{vv}\Lb \omega, v\Rb\, -\,4\, v \,{\cal G}'_v\Lb \omega, v\Rb\,+\,\Bigg\{-2 +\frac{1}{\kappa} \omega- \frac{1 - v^2}{4\,\kappa^2}  \Bigg\}\,{\cal G}\Lb \omega, v\Rb\,\,=\,\,0
\eeq
In Ref.\cite{KOLEV} it is noted, that  function ${\cal G}\Lb \omega, u\Rb$ are intimately related to prolate spheroidal wave functions  $S_{n,m}\Lb c, v\Rb = (1- v^2)^{\frac{m}{2}}\,{\cal G}\Lb c, v\Rb$\cite{ABST,SPHF}, which satisfy the following equation:

\beq \label{GSOL6}
\frac{d}{d v}\Bigg( (1- v^2)\frac{d\,S_{n,m}\Lb c, v\Rb}{d \,v}\Bigg)\,\,+\,\,\Lb \lambda^m_n \,-\,c^2\,v^2 - \frac{m}{1 - v^2}\Rb S_{n,m}\Lb c, v\Rb\,\,=\,\,0
\eeq
with $n$ and $m$ being integer numbers.

For functions  ${\cal G}\Lb c, v\Rb$   \eq{GSOL6} takes  the form:
\beq \label{GSOL7}
(1 - v^2)\, {\cal G}''_{v v}\Lb c, v\Rb\,\,-\,\,2 (m + 1) v \,{\cal G}'_v\Lb c, v\Rb\,\,+\,\,\Big\{ \lambda^m_n\,-\,c^2\,v^2 - m (m+1)\Big\} {\cal G}\Lb c, v\Rb\,\,=\,\,0
\eeq
 Comparing \eq{GSOL5} and \eq{GSOL7} we obtain that
 \beq \label{GSOL8}
  m =1;~~c^2= - \frac{1}{4 \,\kappa^2};~~\omega_n \,\,=\,\,\kappa \,\lambda^1_n + \frac{1}{4\,\kappa};
  \eeq
  Hence, the set of the eigenfunctions for the generating function $Z\Lb \omega, u\Rb$ has the form:
  \beq \label{GSOL9}
  Z_n\Lb v\Rb\,\,=\,\,\, \sqrt{1 - v^2}\,e^{ - \frac{ 1- v}{4 \,\kappa}}\,S_{n,1}\Lb \frac{i}{2 \,\kappa},v\Rb
  \eeq 
  
  Going back to rapidity representation, one can see that
  
    \beq \label{GSOL10}
  Z_n \Lb Y, v\Rb\,\,=\,\,\,e^{ \omega_n\,\Y} \, \sqrt{1 - v^2}\,e^{ - \frac{ 1- v}{4 \,\kappa}}\,S_{n,1}\Lb \frac{i}{2 \,\kappa},v\Rb \,=\,\,e^{ \kappa \lambda^1_n\,\Y}\sqrt{1 - v^2}\,e^{ - \frac{ 1- v}{4 \,\kappa}}\,S_{n,1}\Lb \frac{i}{2 \,\kappa},v\Rb
  \eeq

  Therefore, one can see  that each eigenfunction increases as function of $\Y$  (see Table I for the values of $\omega_n$). 
  \begin{table}
  \begin{tabular}{|c|c|l|l|l|l|l|}
  \hline
  $\omega$ & $\kappa$ & 1&2&3&4&5\\
  \hline
  $\omega_n$ & 8 ($N_c=3$)&1.636&1.766&2.616&3.579&4.8\\
  \hline
  $\omega_n$ & 1 ($N_c=1$)&2.2&6.14&12.133&20.113&30.128\\
  \hline
  \end{tabular}
  \caption{First five eigenvalues $\omega_n $ of  $\omega_n = \kappa\,\lambda^1_n$.}
  \end{table}  
  Generally speaking the generating function is equal to
  \beq \label{GSOL11}
  Z\Lb Y, v\Rb= Z^{\rm asymp} \Lb u\Rb\,+\,\sum_{n=1}^\infty   {\rm C_n} e^{ \omega_n\,\Y} \, \sqrt{1 - v^2}\,e^{ - \frac{ 1- v}{4 \,\kappa}}\,S_{n,1}\Lb \frac{i}{2 \,\kappa},v\Rb 
  \eeq
  where 
  \beq \label{GSOL12}
  {\rm C_n}\,\,=\,\,\int^1_{-1} d v \Bigg( Z\Lb Y=0, u\Rb - Z^{\rm asymp} \Lb u\Rb\Bigg) \,\frac{e^{  \frac{ 1- v}{4 \,\kappa}}  }{ \sqrt{1 - v^2}}\,\,\frac{S_{n,1}\Lb \frac{i}{2 \,\kappa},v\Rb  }{  ||S_{n,1}\Lb \frac{i}{2 \,\kappa},v\Rb||}
  \eeq
   with 
  
  \beq \label{GSOL13} 
  ||S_{n,1}\Lb \frac{i}{2 \,\kappa},v\Rb|| \,\,=\,\,\int^1_{-1} \!\!\!d v\, | S_{n,1}\Lb \frac{i}{2 \,\kappa},v\Rb|^2\,\,=\,\,\frac{2\,n\,(n+1)}{2\,n\,+\,1}
  \eeq 
  
  $Z^{\rm asymp} \Lb u\Rb$ is the solution to \eq{5} which correponds to the minimal value of $\omega =0$. It is convenient to choose it in the following form:
  \beq \label{GSOL14}
    Z^{\rm asymp} \Lb u\Rb\,\,=\,\,\frac{1 -e^{-\frac{u}{\kappa}}}{1 - e^{-\frac{1}{\kappa}} }   
    \eeq
    This solution gives $ Z^{\rm asymp} \Lb u=1\Rb =1$ and $Z^{\rm asymp} \Lb u=0\Rb =0$. Hence  the difference $ Z\Lb Y=0, u\Rb - Z^{\rm asymp} \Lb u\Rb$   satisfies the following boundary condition:
      \beq \label{GSOL15}    
     Z\Lb Y=0, u =0\Rb - Z^{\rm asymp} \Lb u=0\Rb  = 0;~~~       Z\Lb Y=0, u =1\Rb - Z^{\rm asymp} \Lb u=1\Rb = 0;
     \eeq
     This difference can be expanded in the series of \eq{GSOL11} since functions $S_{n,1}\Lb \frac{i}{2 \,\kappa},v\Rb$ satisfy the boundary conditions of \eq{GSOL15} being equal to zero at $v= \pm 1$\footnote{We wish to refer to Refs.\cite{KOLEV,POLY} for more detailed analysis  of solutions to  since \eq{ZEQ2} belong to this class.}  
     
 From \eq{GSOL11} we conclude that $Z\Lb Y, u\Rb$ increases with energy (rapidity Y). In other words even in the simple case of the initial condition of \eq{IC} , which corresponds to the deep inelastic scattering the shadowing corrections  failed to stop the increase of the scattering amplitude.
\subsection{Energy growth from the general structure of  equation.}

  In this section we demonstrate that the energy increase    actually stems from the general structure of \eq{ZEQ2}. Indeed, \eq{GSOL2} can be rewritten in the form of the 
 Sturm-Liouville equation:
 \beq \label{GS1}
  s(u) \,\omega \,Z\Lb \omega, u\Rb + \frac{d}{d\,u} \Bigg( p(u) Z'_u\Lb \omega, u\Rb\Bigg)\,\,=\,\,0 ~~~\mbox{with}~~s(u) =\frac{\kappa}{u (1 - u)}e^{\frac{u}{\kappa}}~~\mbox{and} ~~p(u) = e^{\frac{u}{\kappa}}  \eeq

  The Sturm- Liouville equation has the following general features\cite{KOLEV,POLY}:
  \begin{enumerate}
  \item\quad \eq{GS1} has infinite set of eigenvalues $\omega_m = \lambda_n$.  $\lambda_n$ monotonically increases with $n$ with $\lambda _n \to \infty$  at large $n$. In our case of \eq{GS1} all $\lambda_n > 0$. The least value of  $\lambda_n$ is  $\lambda_0 =0$ which corresponds the asymptotic solution of \eq{ZAS}.
  \item\quad The multiplicity of each eigenvalue is equal to 1.
   \item \quad The eigenfunctions $Z_n(u)$ are orthogonal 
   \beq \label{GS2}
   \int^1_0\!\!
  \!du\,s(u) \,Z_n(u)\,Z_m(u)\,\,=\,0 ~~~\mbox{for}~~n \neq n
   \eeq
  \item \quad For large $n$ 
  \beq \label{GS3}
  \lambda_n\,\,\frac{\pi^2\,n^2}{\delta^2} ~~\mbox{with}~~\delta = \int^1_0 \!\!\!d u \sqrt{\frac{s(u)}{p(u)}}
  \eeq
     \item \quad  For our equation
     \beq \label{GS4}
     \delta =\sqrt{\kappa} \int^1_0 \frac{ d u}{\sqrt{u(1-u)}} \,=\,\,\sqrt{\kappa}\,\pi  
     \eeq
     giving 
     \beq \label{GS5}
     \lambda_n = \,\,\kappa\,n^2
     \eeq
     \end{enumerate}

   Having \eq{GS5} we can conclude, that the generating function $Z_n\Lb \Y,  u \Rb$  
    increases with $Y$. This feature is based on the general features of the Sturm-Liouville equation  and can be  stated  without finding the exact solution.  Since the class of Sturm-Lioville equations is much  wider than our particular equation  (see \eq{ZEQ2}) , we believe that more complicate equations in the case of QCD will still have these property,

\section{Conclusions}


   The main question, that we  have answered in this paper, whether   the shadowing correction can stop the steep increase of  Green's function for the exchange of the $n$ BFKL Pomerons: $G_{n \pom}\Lb Y\Rb \propto \exp\Lb \frac{n^2}{N^2_c} \Delta_{\mbox{\tiny BFKL}} Y\Rb$.  In this paper we considering the simple  Pomeron calculus in zero transverse dimension. This approach has two great advantages: (i)  it  takes into account all shadowing corrections including the summation of the Pomeron loops and (ii) it has the same as in QCD striking increase of $G_{n \pom}\Lb Y\Rb $. Solving exactly the evolution equation we demonstrate that the shadowing corrections cannot stop the increase of scattering amplitude which violates the unitarity constraints. Hence, our answer to the question in the title is positive.
    
   However,
    another phenomenon could considerably increase the shadowing corrections: the saturation effects inside the parton cascade. In the two dimensional Pomeron calculus such corrections have been included  in UTM model \cite{MUSA,BIT,utm,utmm}.  On one hand,  we have demonstrated in Ref.\cite{utmm} the scattering amplitude for  this model in spite of saturation in the parton cascade coincides with the BFKL cascades at high energies.  On the other hand, we need to include $\Gamma^{2 \pom}_{2 \pom}$ in this model. It has not been done and we consider this as a next problem to be solved.
  If we think about the theoretical realization of the parton model (see for example Ref.\cite{WEZU}), we do not expect that the unitarity would be violated. Hence, perhaps all our problems stem from the use of the Pomeron calculus and we need to find the theoretical description beyond this approach.

    The violation of unitarity stems from the vertex $\Gamma^{2 \pom}_{2\pom} >0$ , which also 
    appears in QCD (see Refs.\cite{LRS,BART0,BARTU,LLS,LLR} ) with the same sign. 
   We do not believe that the technical complications, coming with the QCD analysis, could provide the stronger shadowing than in the simple two dimensional models. However, we are aware that we have no idea how to sum Pomeron loops in QCD and how they influence the strength of   the shadowing corrections. On the other hand, we  need to consider a possibility to go out of the Pomeron calculus to treat the high energy  amplitude in CGC (see  for example Ref.\cite{LIEFF}).

   Concluding, we believe that CGC approach  correctly describes the high energy interaction at $N_c\to \infty$ and we do not have a clue how the shadowing correction could suppress the growth of the scattering amplitude in $1/N_c$  order.

~

     {\bf Acknowledgements}

   We thank our colleagues at Tel Aviv university and UTFSM for
 encouraging discussions. Special thanks go A. Kovner, M. Li  and M. Lublinsky for stimulating and encouraging discussions on the subject of this paper. This research    was supported  by 
 ANID PIA/APOYO AFB180002 (Chile), Fondecyt (Chile) grant 1191434 and the Tel Aviv university encouragement grant \#5731.

\end{document}